\def\year{2022}\relax
\documentclass[letterpaper]{article} 
\usepackage{aaai22}  
\usepackage{times}  
\usepackage{helvet}  
\usepackage{courier}  
\usepackage[hyphens]{url}  
\usepackage{graphicx} 
\usepackage{natbib}  
\usepackage{caption} 
\DeclareCaptionStyle{ruled}{labelfont=normalfont,labelsep=colon,strut=off} 
\usepackage{algorithm}
\usepackage{algorithmic}

\usepackage{newfloat}
\usepackage{listings}
\floatstyle{ruled}
\newfloat{listing}{tb}{lst}{}
\floatname{listing}{Listing}
\usepackage{subfig}
\usepackage{amsmath}
\usepackage{makecell}
\usepackage{tabularx}
\usepackage{threeparttable}
\usepackage{multirow}
\usepackage{booktabs}
\usepackage{marvosym}
\usepackage{bm}
\usepackage{mwe}

\newcommand{\ie}{\emph{i.e.,}\xspace}

\newcommand{\etal}{\emph{et al.}\xspace}
\newcommand{\aka}{\emph{a.k.a.,}\xspace}
\newcommand{\wo}{\emph{w / o }\xspace}
\let\oldhat\hat
\renewcommand{\vec}[1]{\mathbf{#1}}
\renewcommand{\hat}[1]{\oldhat{\mathbf{#1}}}
\renewcommand{\matrix}[1]{\mathbf{#1}}

\title{Multi-View Intent Disentangle Graph Networks for Bundle Recommendation}
\author{
    Sen Zhao\textsuperscript{\rm 1,\rm 3}
    Wei Wei\thanks{{} {} Corresponding author.}\textsuperscript{\rm 1,\rm 3},
    Ding Zou\textsuperscript{\rm 1,\rm 3},
    Xianling Mao\textsuperscript{\rm 2}
}
\affiliations{
    \textsuperscript{\rm 1} Cognitive Computing and Intelligent Information Processing (CCIIP) Laboratory, School of Computer Science and Technology, Huazhong University of Science and Technology\\
    \textsuperscript{\rm 2} Beijing Institute of Technology\\
    \textsuperscript{\rm 3} Joint Laboratory of HUST and Pingan Property \& Casualty Research (HPL)\\
    \{senzhao, weiw, m202173662\}@hust.edu.cn, maoxl@bit.edu.cn
}

\usepackage{bibentry}

\begin{document}

\maketitle

\begin{abstract}

Bundle recommendation aims to recommend the user a bundle of items as a whole. Previous models capture the user's preferences on both items and the association of items. Nevertheless, they usually neglect the diversity of the user's intents on adopting items and fail to disentangle the user's intents in representations. In the real scenario of bundle recommendation, a user’s intent may be naturally distributed in the different bundles of that user (Global view), while a bundle may contain multiple intents of a user (Local view).  Each view has its advantages for intent disentangling: 1) From the global view, more items are involved to present each intent, which can demonstrate the user's preference under each intent more clearly. 2) From the local view, it can reveal the association among items under each intent since items within the same bundle are highly correlated to each other. 
To this end, we propose a novel model named \underline{M}ulti-view \underline{I}ntent \underline{D}isentangle \underline{G}raph \underline{N}etworks (MIDGN), which is capable of precisely and comprehensively capturing the diversity of the user's intent and items' associations at the finer granularity.  
Specifically, MIDGN disentangles the user's intents from two different perspectives, respectively: 1) In the global level, MIDGN disentangles the user's intent coupled with inter-bundle items; 2) In the Local level, MIDGN disentangles the user's intent coupled with items within each bundle.
 Meanwhile, we compare the user's intents disentangled from different views under the contrast learning framework to improve the learned intents.  Extensive experiments conducted on two benchmark datasets demonstrate that MIDGN outperforms the state-of-the-art methods by over 10.7$\%$ and 26.8$\%$, respectively. The implementation of our proposed models is publicly available at https://github.com/Snnzhao/MIDGN.git.
\end{abstract}
\section{ Introduction}
 Recently, the blossom of Internet Technology has revolutionized our routine activities by providing a tremendous amount of valuable contents, which however lead to the serious information overload problem\cite{zhao2019recurrent, wei2019emotion}.  To alleviate this problem, recommendation system \cite{zheng2010collaborative} has been an effective tool to help the user quickly discover items of potential interest to themselves. In general, conventional recommendation systems \cite{wu2019session, wang2020disenhan, wang2020global,wang2020exploiting,wang2020exploring} aim to recommend individual items to users, but in practice (e.g., E-commerce), items bundling is also a widely-adopted solution to increase the exposure of items that are seldom purchased in isolation. To this end, the problem of bundle recommendation (BR)\cite{zhu2014bundle} is proposed to recommend a bundle of items that the user has interest to purchase together. 
\begin{figure}[!t]
    \centering\includegraphics[width=0.45\textwidth]{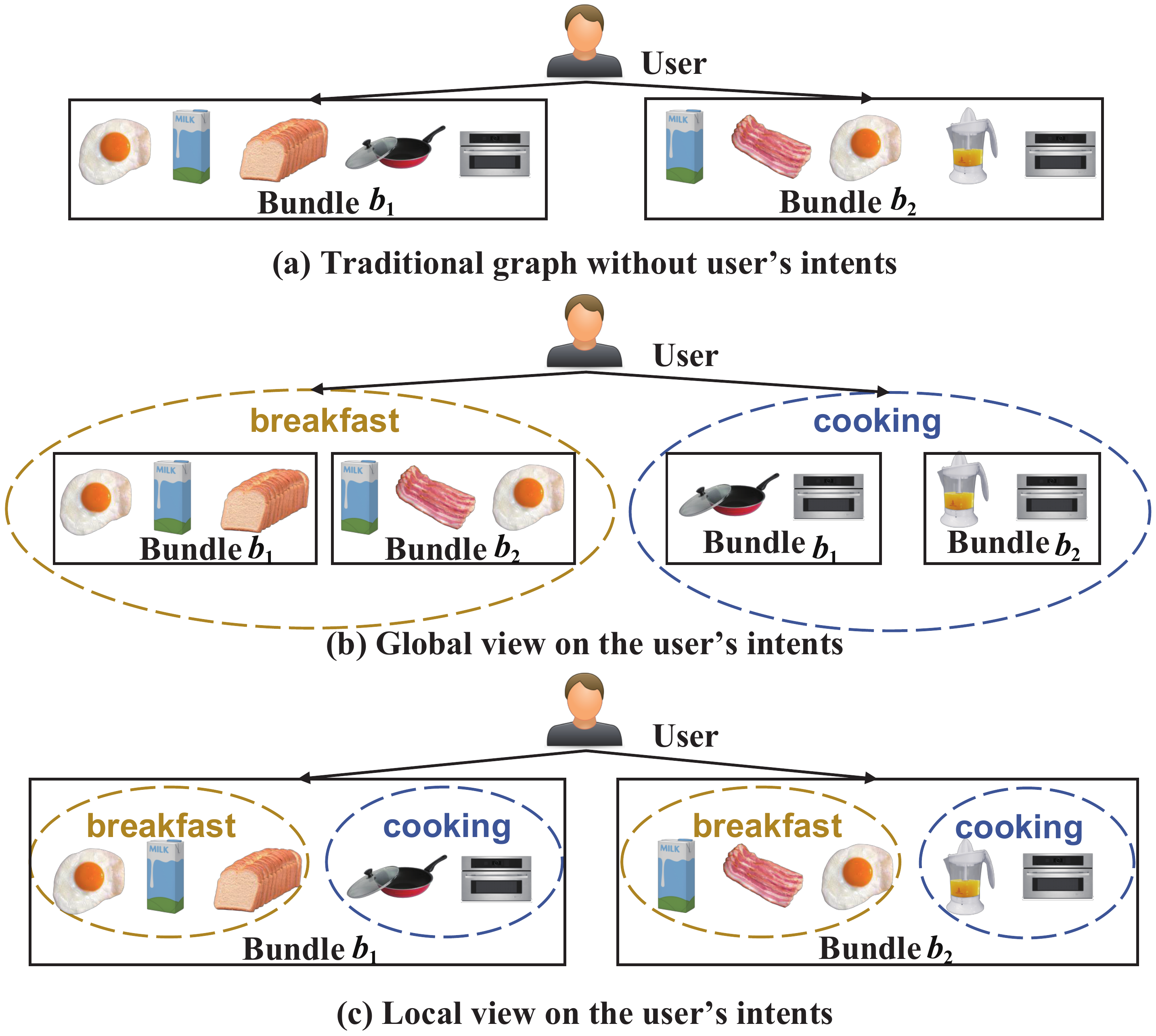}
    \caption{An example of the user's multi-intent pattern for bundle recommendation.}
    \label{fig:intuition}
  \end{figure}
  
Similar to conventional recommendation system, learning the user's preferences is the main theme for bundle recommendation. Existing methods mainly focus on capturing the user's preferences on both items and the bundle of items. Liu \etal \shortcite{liu2014recommending} and Cao \etal \shortcite{cao2017embedding} simultaneously utilize the user’s interactions with both items and bundles under the Bayesian Personalized Ranking(BPR) \cite{rendle2014bayesian} framework. Chen \etal \shortcite{chen2019matching} jointly model user-bundle and user-item interactions in a multi-task manner, but they may neglect the user's preference over the association of items within bundles, which is also of crucial importance to bundle recommendation.  Chang \etal \shortcite{chang2020bundle} unify user-bundle-item affiliation into a heterogeneous graph and uses graph convolution networks (GCN) to learn the  user and bundle's representation, capturing item level semantics, however these works may neglect the diversity of the user's intents on adopting items and fail to disentangle the user's intents in representations. Wang \etal \shortcite{wang2020disentangled} consider user-item relationships at the finer granularity of the user's intents and generate disentangled representations for the item recommendation problem. Nevertheless, the real scenarios of bundle recommendation are more complicated (as shown in Figure \ref{fig:intuition}), a user’s intent may be naturally distributed in the different bundles of that user. (\emph{Global view}), while a bundle may contain multiple intents of a user (\emph{Local view}). 
   There are two obvious drawbacks when neglecting the diversity of the user's intents: 1) from the global view, the inter-bundle items under different intents can interfere with each other. As illustrated in Figure \ref{fig:intuition}(b), the egg can be noisy for constructing the intent of cooking. 2) from the local view, the association of inter-intent items may have a negative rating on the user's interest for each bundle. For the example in Figure \ref{fig:intuition}(b), bundle $b_1$ may not be recommended because bread and pan are rarely purchased together. Each view has its own advantages for intent disentangling: 1) In the global view, more items are involved to present each intent, which can demonstrate the user's preference under each intent more clearly. 2) The local view can reveal the association among items under the user’s each intent since the items within the same bundle are highly correlated to each other. Disentangling the user's intent from both views, we can better represent the user's preference and the association among items at a finer grain of the user's intent.

In this work, we propose a novel model, \underline{M}ulti-view \underline{I}ntents \underline{D}isentangle \underline{G}raph \underline{N}etworks (MIDGN), to disentangle the user and bundle representations at the granularity of the user 's intents. Specifically, we disentangle the representations of the user and the bundle into chunks, each chunk represents a latent intent. A graph neural network equipped with neighbour routing is applied both to disentangle the user-item and the bundle-item graph, where such two types of graphs can provide different views of users' intent for recommendation. Furthermore, a contrast learning module comparing the user's intents under different views is also employed for enhancing the learning of user intent representations. Extensive experiments conducted on NetEase and Youshu demonstrate that our proposed model MIDGN outperforms the state-of-the-art methods by over 10$\%$.

In a nutshell, this work makes the following contributions:
\begin{itemize}
    \item We emphasize the importance of disentangling the user 's intent in bundle recommendation problem and explore two views for the user 's intents disentangling.

    \item We propose the model MIDGN, which disentangles the user and bundle's representation by modeling and contrasting the user's intents under global and local views.

    \item We conduct extensive experiments on two benchmark datasets and MIDGN achieves over 10 percents improvement over the state-of-art methods.
\end{itemize}
\section{ Related Work}
The early work \cite{chen2005market,garfinkel2006design} on bundle recommendation mainly focus on mining a group of common items from historical user-item interactions for users to consume together, while ignoring the relevance of the selected items. For example, \cite{liu2011personalized, xie2010breaking} formalize the bundle recommendation task as a (linear) knapsack problem and no item-to-item dependency is considered. \cite{parameswaran2011recommendation} models the curriculum recommendation as a \emph{constraint satisfaction problem} and make use of cross-item dependencies as general hard-constrains for recommendation. However, previous bundle recommendation methods are incapable of discovering the items of common interest, and thus cannot accurately capture  a use's personalized preference for recommendation.

In recent years, many studies have been devoted for bundle recommendation, which can be roughly categorized into two classes, (1) Intra-basket recommendation \cite{le2017basket}, which recommends items to be added to the current basket. 
There exist several attempts on this problem, for example, Le \etal \shortcite{le2017basket} propose a factorization based model (named CBFM) to  identify the correlations of items within basket to integrate constraints for the given basket with similar intent. 
Liu \etal \shortcite{liu2017modeling} employ a probabilistic graphical model to infer users' preferences over bundle via introducing two latent factors, \ie node-type (items) and edge-type (the associations between pairs of items). (2) Inter-basket (\aka next-basket), which offers multiple items for sale as a bundle (\aka bundle list recommendation \cite{sar2016beyond,bai2019personalized}).
Generally, bundle recommendation is formalized as a top-K optimization problem, which can be solved with learning-to-rank based approaches. For instance, Xie \etal \shortcite{xie2014generating} take account of sampling and preference elicitation based method to generate top-K bundles for users, in which an additive utility function with a combination of the corresponding  feature values is employed for tackling the problem, but may fail to capture the complicated dependencies of user-item-bundle interactions. Further efforts are conducted to capture the complicated dependencies. 
 Cao \etal \shortcite{cao2017embedding} simultaneously utilize the users’ interactions with both items and bundles under the Bayesian Personalized Ranking(BPR) \cite{rendle2014bayesian} framework. Chen \etal \shortcite{chen2019matching} jointly model user-bundle and user-item interactions in a multi-task manner. These methods, however, fail to learn multi-hop connectivity from user-bundle-item interaction  and neglect user’s preference over the association of items within bundles which is also crucial for bundle recommendation.

Recently, graph convolution networks (GCN) \cite{kipf2016semi} has been extensively used for recommendation due to its superior ability in learning of the graph structure.
Berg \etal \shortcite{berg2017graph} first applie GCN to recommendation to factorize several rating matrices.  Wang \etal \shortcite{wang2019neural} employ GCN to encode the collaborative signal in the form of high-order connectivities. Graph convolution networks are also adopted to solve the problem of bundle recommendation.  Liu \etal \shortcite{liu2020basconv} designs three types of aggregators
with GCN specifically for users, bundles and items. Chang \etal \shortcite{chang2020bundle} unify user-bundle-item affiliation into a heterogeneous graph and uses GCN to learn user and bundle representation capturing item level semantics. These works, however, neglect the diversity of user's intents on adopting items and fail to disentangle user's intents in representations.

Disentangling user's intents has been a topic in the problem of item recommendation. Wang \etal \shortcite{wang2020disentangled} considers user-item relationships at the finer granularity of user's intents and generates disentangled representations. But for the problem of bundle recommendation, where the user's intents are distributed inter-and intra-bundles, disentangling user's intents is more complicated and new methods should be studied.
\section{Preliminaries}
Let $\mathcal{U}=\{u_1, u_2,\cdots, u_M\}$ be the user set, and $\mathcal{B}=\{b_1, b_2,\cdots, b_O\}$ and  $\mathcal{I}=\{i_1, i_2,\cdots, i_N\}$ be the associated bundle set and item set, where M, O, N denote the number of users, bundles and items. 
 According to the history of bundles user consumed, we can define the user-bundle interaction matrix, bundle-item affiliation matrix and user-item interaction matrix as $\matrix{Y}_{M\times O}=\{ y_{ub}|u\in\mathcal{U}, b\in\matrix{B}\}$, $\matrix{H}_{O\times N}=\{ h_{bi}|b\in\mathcal{B}, i\in\matrix{I}\}$ and $\matrix{R}_{M\times N}=\{ r_{ui}|u\in\mathcal{U},   i\in\mathcal{I}\}$, in which $y_{ub}=1$, $h_{bi}=1$ and $r_{ui}=1$ means user $u$ once interacted bundle $b$, bundle $b$ contains item $i$, and user $u$ once bought item $i$, respectively.

The task of bundle recommendation is to predict the probability of user $u$ potentially interacting with a given bundle never seen before. Specifically, our goal is to learn a predict function $\hat{y}=\mathcal{F}(u,b|\theta, \matrix{H})$, where $\hat{y}$ is the estimation probability and $\theta$ implies the parameters of function $\mathcal{F}$.
\section{The Proposed Model}
\begin{figure*}[!ht]
    \centering\includegraphics[width=0.95\textwidth]{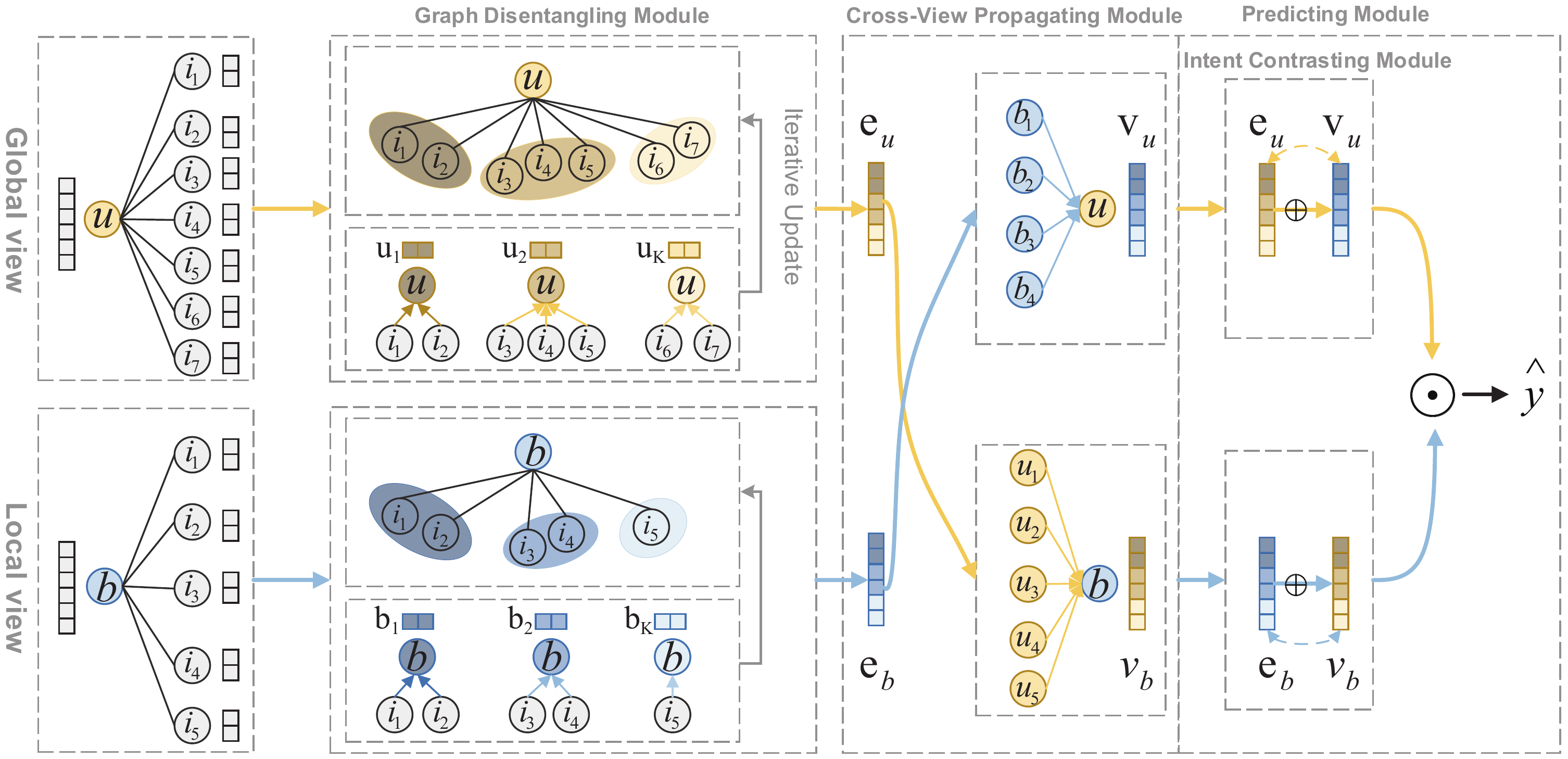}
    \caption{The framework of the proposed MIDGN. It contains four modules: (a) \emph{Graph disentangling module}, which disentangles user-item   and bundle-item interactions coupling with the user's intents under global and local view respectively; (b) \emph{Cross-view propagating module} that propagates collaborative signal coupling with the user's intents under different views; (c) \emph{Intent contrasting module} to  encourage the correlation of the user’s intents under different views and \emph{Predicting module}. (Best view in color)}
    \label{fig:structure}
  \end{figure*}

In this section, we present Multi-view Intent Disentangle Graph Networks (MIDGN)(shown in Figure \ref{fig:structure}), which is composed of four different components: 1) \emph{Gaph disentangling module}, which disentangles user-item  interactions and bundle-item interactions coupling with the user's intents from \emph{global} and \emph{local} perspectives, respectively; 2) \emph{Cosss-view propagating module} that propagates collaborative signal coupling with the user's intents under different views through user-bundle interactions; 3) \emph{Intent contrasting module}, which employs InfoNCE \cite{oord2018representation} to encourage the correlation of the user's intents under different views; and \emph{predicting module}, which predicts the rating with the learned user and bundle embeddings.
\subsection{Graph Disentangling Module}
In this module, we first slice each user/bundle embedding into $K$ chunks and couple each chunk with the user's one intent. Then a graph neural networks incorporated with neighbor routing mechanism is devised to disentangle the user-item/bundle-item graph and refine the intent-aware user/bundle representation. 

\textbf{Initialization of intent-aware embeddings and graphs.} We assume the user has $K$ intents and slice each user/bundle's embedding into $K$ chunks in different feature space.  Coupling with each intent, a chunk of the user and bundle's embedding is independently initialized respectively. Formally, the user and bundle's embedding are represented as:
\begin{equation}
\begin{split}
 \vec{u}=(\vec{u}_{1}, \vec{u}_{2}, \cdots, \vec{u}_{K}),\\
 \vec{b}=(\vec{b}_{1}, \vec{b}_{2}, \cdots, \vec{b}_{K}),
\end{split}
\label{eq:embedding}
\end{equation}
where $\vec{u} \in R^{d}$ and $\vec{b} \in R^{d}$ are embedding of the user and bundle.  $\vec{u}_{k}\in R^{\frac{d}{K}}$, $\vec{b}_k\in R^{\frac{d}{K}}$ are the embedding chunks of the user and bundle under the $k$-th intent, respectively. 
As each item is bought by the user under one intent, there is no need to divide it's embedding. So the item embedding is presented as $\vec{i} \in R^{\frac{d}{K}}$ and randomly initialized.

We assume each intent $k$ is likely to be the reason why the user/bundle connects with a certain subset of its neighbour items.
To present user/bundle-item interactions motivated by different intents, a set of intent-aware graphs $\mathcal{G}=\{\mathcal{G}_1, \mathcal{G}_2, \cdots, \mathcal{G}_K\}$ is built. For each intent-aware graph $\mathcal{G}_k$, a weighted adjacent matrix $\matrix{A}_k$ is built, where $\matrix{A}_k(c, i)$ denotes the confidence that the interaction between $c$ and item $i$ is motivated by the $k$-th intent. Here $c$\footnote{This paper uses symbol c as a uniform placeholder for user symbol $u$ and bundle symbol $b$, since the calculation and formulation for the user and bundle in this article are similar in many scenarios} is a uniform placeholder for user symbol $u$ and bundle symbol $b$. Intuitively, each interaction within user/bundle-item graph is associated with the user's $K$ intents to different degree as follows:
\begin{equation}
 \matrix{A}(c, i)=(\matrix{A}_1(c, i), \matrix{A}_2(c, i), \cdots, \matrix{A}_K(c, i)),
\label{eq:graph}
\end{equation}

 We uniformly initialize the confidence of each interaction under the user's intents as:
 \begin{equation}
 \matrix{A}(c, i)=(1, 1, \cdots, 1),
\label{eq:initial_graph}
\end{equation}
which denotes intents contributes equally to the each interaction. 

\textbf{Intent-aware interaction graph disentangling.} Each intent $k$ is specialized with embedding chunk $\vec{c_k}$ as well as a specific interaction graph $\mathcal{G}_k$. When describe user/bundle $c$ under intent $k$, we should construct $\vec{c_k}$ with neighbours that connect to user/bundle $c$ due to intent $k$. To realize this, a graph disentangling layer is employed to user-item and bundle-item graph as follows:
\begin{equation}
    \vec{e}_{ck}^{(1)}=g(\vec{c}_k, \{\vec{i}, \vec{i}\in \mathcal{N}_c\}),
\label{eq:disentangle0}
\end{equation}
where $g(\cdot)$ denotes the graph disentangling layer. $i \in \mathcal{N}_c$ denotes items that have interaction with user/bundle $c$. $\vec{e}_{ck}^{(1)}$ is to capture the collaborative signal related to the user's $k_{th}$ intent within node $c$'s neighbor. The super-index (1) denotes the first-order neighbors.

As illustrated in Figure \ref{fig:structure}, the neighbor routing mechanism is adopted in the graph disentangling module, which iteratively updates the user/bundle's embedding chunks $\vec{c}_k$ and the adjacent matrix $\matrix{A}_k$ for graph $\mathcal{G}_k$. For each iteration, $\vec{c}_k^t$  and $\matrix{A}_k^t$ are used to memorize the update of $\vec{c}_k$ and $\matrix{A}_k$.

\textbf{Update within each iteration.} For each interaction $(c, i)$, we have its confidence under $K$ intents $\{\matrix{A}_k^t(c, i) | \forall k \in \{1, 2,\cdots,K\}\}$ . To calculate it's distribution over intents, softmax function is applied to the confidence: 
\begin{equation}
    \tilde{\matrix{A}}_k^t(c, i)=\frac{exp{\matrix{A}}_k^t(c, i)}{{\sum_{k'=1}^K} exp{\matrix{A}}_{k'}^t(c, i)},
\label{eq:disentangle2}
\end{equation}
which indicates the extent to which interaction $(c, i)$ is motivated by each intent.

Then the embedding propagation is employed to each intent-aware graph  $\mathcal{G}_k$, to update node $c$'s embedding chunk under different intents with its neighbor items. Specifically, the weighted aggregation is defined as:
\begin{equation}
    \vec{c}_k^t=\sum\limits_{i\in \mathcal{N}_c}\frac{\tilde{\matrix{A}}_k^t(c, i)}{\sqrt{D_t^k(c)\cdot D_t^k(i)}}\cdot \vec{i},
\label{eq:disentangle_ui3}
\end{equation}
where $D_t^k(c)=\sum\limits_{i \in \mathcal{N}_c}{\tilde{\matrix{A}}_k^t(c, i)}$ and $D_t^k(i)=\sum\limits_{c \in \mathcal{N}_i}{\tilde{\matrix{A}}_k^t(c, i)}$ are degrees of node $c$ and  item  respectively. This normalization is to deal with the influence of varying numbers of nodes on the training process.

Then for each node $c$, we should update its confidence of connections with items $\matrix{A}^t_k(c, i)$.  To realize this, the confidence  $\matrix{A}^t_k(c, i)$ of interaction $(c, i)$ under each intent  is updated through: 
\begin{equation}
    \matrix{A}_k^{t+1}(c, i)=\matrix{A}_k^t(c, i)+{\vec{c}_k^t}^\mathrm{T}\cdot \vec{i}.
\label{eq:disentangle5}
\end{equation}
If the item's embedding $\vec{i}$ are similar with the node $c$'s embedding chunk ${\vec{c}_k^t}$ under intent $k$, the confidence of their connection $\matrix{A}_k(c, i)$ will be enhanced.

\textbf{Layer combination.} $\vec{e}_{ck}^1$ involve one-hop information within node $c$'s neighbour. To incorporate multi-hop, more disentangle graph layer  is employed as:
\begin{equation}
    \vec{e}_{ck}^l=g(\vec{e}_{ck}^{l-1}, \{\vec{i}, \vec{i}\in \mathcal{N}_c\}).
\label{eq:disentangle6}
\end{equation}
The intent-aware representation from different layers is summed to get the final representation:
\begin{equation}
    \vec{e}_{ck}=\sum_{l}\vec{e}_{ck}^l
\label{eq:disentangle_ui7}
\end{equation}

The graph disentangling module learns the user's intents distributed in different bundles (global view) from user-item graph. And from bundle-item graph, the graph disentangling module learns the user's multiple intents within each bundle (local view). Coupling with intents from global and local views, we disentangle the representation  of user and bundle, respectively. Specifically, this module disentangles the representation of user and bundle into chunks as $\vec{e}_u=(\vec{e}_{u1}, \vec{e}_{u2}, \cdots, \vec{e}_{uK})$
and $\vec{e}_b=(\vec{e}_{b1}, \vec{e}_{b2}, \cdots, \vec{e}_{bK})$.
\subsection{Cross-View Propagating Module}
The user's representation $\vec{e}_u$ and bundle's  representation $\vec{e}_b$ couple with the user's intents from global view and local view, respectively. To exchange intents under different views between the chunks of user and bundle, the LightGCN \cite{he2020lightgcn} is employed to the user-bundle graph:
\begin{equation}
\begin{aligned}
    \vec{v}_u=\sum_{b\in \mathcal{N}_u}\frac{1}{\sqrt{|\mathcal{N}_u|}\sqrt{|\mathcal{N}_b|}}\vec{e}_b,\\
    \vec{v}_b=\sum_{u\in \mathcal{N}_b}\frac{1}{\sqrt{|\mathcal{N}_b|}\sqrt{|\mathcal{N}_u|}}\vec{e}_u.
\end{aligned}
\label{eq:aggregator1}
\end{equation}
Through exchanging views, the chunks $\vec{v}_u$ and $\vec{v}_b$ represent the user and bundle coupling with intents under local and global view respectively.

\subsection{Intent Contrasting Module}
From the disentangling module and cross-view propagating module, we obtain the user and bundle's embedding chunks $\vec{e}_c=(\vec{e}_{c1},\cdots,\vec{e}_{cK})$ and $\vec{v}_c=(\vec{v}_{c1},\cdots,\vec{v}_{cK})$ coupling with intents from global and local view, respectively. In the global view, more items are involved to present each intent, which can demonstrates user's preference under each intent more clearly. The local view can reveal the association among items under the user’s each intent. By contrasting, intents from different views complement each other and can better present the user's preference as well as items' associations at a finer grain of the user's intent. Hence we introduce another module which applies InfoNCE\cite{oord2018representation} to contrast the user and bundle's embedding capturing intents from different views:
\begin{equation}
    L_{contrast}=-log(\frac{exp(\vec{e}_{ck} \cdot \vec{v}_{ck_+})}{\sum_{k'}exp(\vec{e}_{ck} \cdot \vec{v}_{ck'})}),
\label{eq:contrast2}
\end{equation}
where for each chunk of user/bundle's embedding $\vec{e}_{ck}$, the positive sample $\vec{v}_{ck_+}$ is the chunk of the same intent within embedding from the different view, while all other chunks are the negative samples.

\subsection{Prediction and Optimization}
Having obtained the presentation of the user and bundle from both global and local views, we concatenate representations from two views to get the final representation and use inner product to estimate the likelihood of their interactions, $\hat{y}_{ub}$ as:
\begin{equation}
    \hat{y}_{ub}=(\vec{e}_u \oplus \vec{v}_u)\odot(\vec{e}_b \oplus \vec{v}_b),
\label{eq:predictionn}
\end{equation}
where $\oplus$ denotes the concatenation and $\odot$ denotes dot product.
Then for model optimization, we adopt the Bayesian Personalized Ranking loss \cite{rendle2012bpr}:
\begin{equation}
    L_{pred}=\sum_{(u, b, d)\in Q}{-\ln\sigma{(\hat{y}_{ub}-\hat{y}_{ud})}}+\lambda\cdot{||\theta||}^2,
\label{eq:optimization}
\end{equation}
where  $Q=\{(u, b, d)| (u, b)\in y^+, (u, d)\in y^-\}$ denotes the training data that involves observed interaction $y^+$ and unobserved interaction $y^-$. To avoid over-fitting, we adopt $L_2$ regularization on the model's parameter $\theta$ which is controlled by the coefficient $\lambda$. During training, we alternatively optimize the prediction loss in eq. (\ref{eq:predictionn}) and contrast loss in eq. (\ref{eq:contrast2}).

\section{ Experiments}
To answer the following questions, we conduct experiments on two public datasets.
\begin{itemize}
    \item \textbf{RQ1:} How does MIDGN perform compared with previous approaches?
    \item \textbf{RQ2:} How do different components (Graph Disentangling module in global view, Graph Disentangling module in local view, intent contrast module) affect the results of MIDGN?
    \item \textbf{RQ3:} How do parameters (layer number of Graph Disentangling module, number of intents) influence the results of MIDGN?
\end{itemize}
{
\begin{table}[h]
    \setlength{\tabcolsep}{5pt}
 \centering
 \small
    \begin{tabularx}{0.45\textwidth}{p{3cm}|X|X}
    \toprule
    \makecell[c]{\text{Dataset}}&\makecell[c]{\text{NetEase}}&\makecell[c]{\text{Youshu}}\cr
    \hline
    \hline
    \makecell[c]{\text{User}}&\makecell[c]{18,528}&\makecell[c]{8,039}\cr

    \makecell[c]{\text{Bundle}}&\makecell[c]{22,864}&\makecell[c]{4,771}\cr
    \makecell[c]{\text{Item}}&\makecell[c]{123,628}&\makecell[c]{32,770}\cr
    \hline
    \makecell[c]{\text{User-Bundle}}&\makecell[c]{302,303}&\makecell[c]{51,337}\cr
    \makecell[c]{\text{Bundle-item}}&\makecell[c]{1,778,838}& \makecell[c]{176,667}\cr
    \makecell[c]{\text{User-item}}&\makecell[c]{1,128,065}&\makecell[c]{138,515}\cr
    \hline
    \makecell[c]{\text{User-bundle density}}&\makecell[c]{0.07$\%$} &\makecell[c]{0.13$\%$}\cr
    \makecell[c]{\text{Bundle-item density}}&\makecell[c]{0.06$\%$} &\makecell[c]{0.11$\%$}\cr
    \makecell[c]{\text{User-item density}}&\makecell[c]{0.05$\%$}&\makecell[c]{0.05$\%$}\cr
    \bottomrule[0.8pt]
    \end{tabularx}
    \caption{Statistics of two utilized datasets}
    \label{tab:data}
    
\end{table}
}
\textbf{Datasets and metrics}
Two datasets are used to evaluate our proposed method. The one is NetEase provided by the work\cite{cao2017embedding}. NetEase is constructed by crawling data from Netease Cloud Music, which enables users to select preferred user-generated playlists or individual songs. Within NetEase, each bundle contains at least 10 songs appearing at least in 5 bundles. Each user consumes at least 10 songs and 10 playlists. The other dataset is named Youshu provided by the work \cite{chen2019matching}. Youshu is constructed by crawling data from Youshu, a Chinese book review site. Youshu contains the user's preference for both individual books and bundles of books. The statistic of these two datasets is shown in Table \ref{tab:data}.

We select Recall@K and NDCG@K as evaluation metrics to judge the ranking list performance. Recall@K  means the ratio of test bundles within the top-K ranking list. NDCG@K assigning higher scores to the hits at a higher position on the ranking list.

\textbf{Baselines} To demonstrate the effectiveness of the proposed MIDGN, state-of-art methods are chosen for comparison. Specifically, we first chose two matrix factorization based methods, and several graph based methods as follows: 
{
\begin{table*}[ht]\scriptsize
  \setlength{\tabcolsep}{2.5pt}
  \centering
  \resizebox{0.9\textwidth}{!}
  {
    \begin{tabular}{c|lcclcclccl|lcclcclcc}
    \toprule
    \multirow{3}{*}{\text{Method}}&&\multicolumn{8}{c}{\text{NetEase}}&&&\multicolumn{8}{c}{\text{Youshu}}\\
    &&\multicolumn{2}{c}{Metrics@20}&&\multicolumn{2}{c}{Metrics@40}&&\multicolumn{2}{c}{Metrics@80}&&&\multicolumn{2}{c}{Metrics@20}&&\multicolumn{2}{c}{Metrics@40}&&\multicolumn{2}{c}{Metrics@80}\cr
    \cline{3-4} \cline{6-7} \cline{9-10} \cline{13-14} \cline{16-17} \cline{19-20} &&\text{Recall}&\text{NDCG}&&\text{Recall}&\text{NDCG}&&\text{Recall}&\text{NDCG}&&&\text{Recall}&\text{NDCG}&&\text{Recall}&\text{NDCG}&&\text{Recall}&\text{NDCG}\cr
     \hline
    MFBPR&&	\makecell[c]{0.0335}&	\makecell[c]{0.0181}&&	\makecell[c]{0.0600}&	\makecell[c]{0.0246}&&	\makecell[c]{0.0948}&	\makecell[c]{0.0323}&&&	\makecell[c]{0.1959}&	\makecell[c]{0.1117}&&	\makecell[c]{0.2735}&	\makecell[c]{0.1320}&&	\makecell[c]{0.3710}&	\makecell[c]{0.1543}\cr
    DAM&&	\makecell[c]{0.0411}&	\makecell[c]{0.0210}&&	\makecell[c]{0.0690}&	\makecell[c]{0.0281}&&	\makecell[c]{0.1090}&	\makecell[c]{0.0372}&&&	\makecell[c]{0.2082}&	\makecell[c]{0.1198}&&	\makecell[c]{0.2890}&	\makecell[c]{0.1418}&&	\makecell[c]{0.3915}&	\makecell[c]{0.1658}\cr
    \hline
    GCNBG&&	\makecell[c]{0.0378}&	\makecell[c]{0.0257}&&	\makecell[c]{0.0625}&	\makecell[c]{0.0255}&&	\makecell[c]{0.1000}&	\makecell[c]{0.0342}&&&   \makecell[c]{0.1982}&	\makecell[c]{0.1141}&&	\makecell[c]{0.2661}&	\makecell[c]{0.1322}&&	\makecell[c]{0.3633}&	\makecell[c]{0.1541}\cr
    NGCFBG&&	\makecell[c]{0.0395}&	\makecell[c]{0.0207}&&	\makecell[c]{0.0646}&	\makecell[c]{0.0274}&&	\makecell[c]{0.1021}&	\makecell[c]{0.0359}&&&	\makecell[c]{0.1985}&	\makecell[c]{0.1143}&&	\makecell[c]{0.2658}&	\makecell[c]{0.1324}&&	\makecell[c]{0.3542}&	\makecell[c]{0.1524}\cr
    \hline
    GCN&&	\makecell[c]{0.0402}&	\makecell[c]{0.0204}&&	\makecell[c]{0.0657}&	\makecell[c]{0.0272}&&	\makecell[c]{0.1051}	&\makecell[c]{0.0362}&&&	\makecell[c]{0.2032}&	\makecell[c]{0.1175}&&	\makecell[c]{0.2770}&	\makecell[c]{0.1371}&&	\makecell[c]{0.3804}&	\makecell[c]{0.1605}\cr
    NGCF&&	\makecell[c]{0.0384}&	\makecell[c]{0.0198}&&	\makecell[c]{0.0636}&	\makecell[c]{0.0266}&&	\makecell[c]{0.1015}&	\makecell[c]{0.0350}&&&	\makecell[c]{0.2119}&	\makecell[c]{0.1165}&&	\makecell[c]{0.2761}&	\makecell[c]{0.1343}&&	\makecell[c]{0.3743}&	\makecell[c]{0.1561}\cr

    RGCN&&	\makecell[c]{0.0407}&	\makecell[c]{0.0210}&&	\makecell[c]{0.0670}& 	\makecell[c]{0.0280}&& 	\makecell[c]{0.1112}&	\makecell[c]{0.0378}&&&	\makecell[c]{0.2040}&	\makecell[c]{0.1069}&&	\makecell[c]{0.3017}&	\makecell[c]{0.1330}&&	\makecell[c]{0.4169}& \makecell[c]{0.1595}\cr
    BasConv&&	\makecell[c]{0.0470}& 	\makecell[c]{0.0244}&&	\makecell[c]{0.0786}&	\makecell[c]{0.0327}&&	\makecell[c]{0.1241}&	\makecell[c]{0.0429}&&&	\makecell[c]{0.2121}&	\makecell[c]{0.1204}&&	\makecell[c]{0.2978}&	\makecell[c]{0.1434}&&	\makecell[c]{0.3988}&	\makecell[c]{0.1548}\cr
   BGCN&&	\makecell[c]{\underline{0.0491}}&	\makecell[c]{\underline{0.0258}}&&	\makecell[c]{\underline{0.0829}}&	\makecell[c]{\underline{0.0346}}&&	\makecell[c]{\underline{0.1304}}&	\makecell[c]{\underline{0.0453}}&&&	\makecell[c]{\underline{0.2347}}&	\makecell[c]{\underline{0.1345}}&&	\makecell[c]{\underline{0.3248}}&	\makecell[c]{\underline{0.1593}}&&	\makecell[c]{\underline{0.4355}}&	\makecell[c]{\underline{0.1851}}\cr
   
    \hline
   \textbf{MIDGN}&& 	\makecell[c]{\textbf{0.0678}}&	\makecell[c]{\textbf{0.0343}}&&	\makecell[c]{\textbf{0.1085}}&	\makecell[c]{\textbf{0.0451}}&&	\makecell[c]{\textbf{0.1654}}&	\makecell[c]{\textbf{0.0578}}&&&	\makecell[c]{\textbf{0.2682}}&	\makecell[c]{\textbf{0.1527}}&&	\makecell[c]{\textbf{0.3712}}&	\makecell[c]{\textbf{0.1808}}&&	\makecell[c]{\textbf{0.4817}}&	\makecell[c]{\textbf{0.2063}}\cr
    $\%$Improv.&& \makecell[c]{38.2$\%$}&	\makecell[c]{33.1$\%$}&&	\makecell[c]{30.9$\%$}&	\makecell[c]{30.3$\%$}&&	\makecell[c]{26.8$\%$}&	\makecell[c]{27.5$\%$}&&&	\makecell[c]{14.2$\%$}&	\makecell[c]{13.8$\%$}&&	\makecell[c]{14.2$\%$}&	\makecell[c]{13.6$\%$}&&	\makecell[c]{10.7$\%$}&	\makecell[c]{11.3$\%$}\cr
   
    \bottomrule
    \end{tabular}
    }
    \caption{Performance comparisons on two real-world datasets}
    \label{tab1}
\end{table*}
}
    \begin{itemize}
    \item \textbf{MFBPR} \cite{rendle2009bpr}: This work applies a Bayesian Personalized Ranking learning framework to the matrix factorization method.
    
    \item \textbf{DAM} \cite{chen2019matching}: This work uses the factorized attention mechanism and multi-task framework to capture bundle-level association and collaborate signals.
    
    \item
    \textbf{GCN}\cite{berg2017graph}: The GCN method is applied to the user-bundle-item unified graph to predict the user-bundle relations.
    \item \textbf{GCNBG}\cite{berg2017graph}:  The GCN method is applied to the user-bundle graph.

    \item \textbf{NGCF} \cite{wang2019neural}:  This work uses GCN based method to capture high-order collaborative signals for prediction. 
    
    \item \textbf{NGCFBG} \cite{wang2019neural}: The NGCF method is applied to the user-bundle graph for prediction.
    
    \item \textbf{RGCN} \cite{schlichtkrull2018modeling}: RGCN is GCN based method developed to deal with the multi-relational graph.

    \item \textbf{BasConv} \cite{liu2020basconv}: This work devises heterogeneous aggregators to learn the embedding of each node within the user-bundle-item graph.
    
    \item \textbf{BGCN} \cite{chang2020bundle}: BGCN proposes a graph neural network model to explicitly model complex relations between users, items, and bundles.
    \end{itemize}
    The negative sampling is set to 1, the learning rate is selected from {1e-5, 3e-5, 1e-4, 3e-4, 1e-3, 3e-3}, and we adopt BPR loss for all methods and employ Adam optimizer with the 4096-size mini-batch and fit the embedding size as 64. We make use of Nvidia Titan RTX graphics card equipped with AMD r9-5900x CPU (32GB Memory). For MIDGN, the number of intents and the number of layers are selected from {1, 2, 4, 8} and {1, 2, 3, 4}, respectively

\subsection{Performance Comparison(RQ1)}
The results of all the methods are reported in Table \ref{tab1}. For the results, we have the following observations.

\begin{itemize}
    \item \textbf{Our proposed MIDGN achieves the best results.} MIDGN significantly outperforms all the baselines both on the metrics of Recall@K and NDCG@K. On NetEase, MIDGN improves the performance over the best baseline by 26.8$\%$-38.2$\%$. And MIDGN outperforms the best result by 10.7$\%$-14.2$\%$ on Youshu. We contribute the improvement to the following aspects:1) By grouping the items to model the user's intents, MIDGN reduces the noise from individual items and thus better learns the user's preference. 2) Disentangling the embedding of user and bundle according to the user's intents presents the user and bundle at a more granular level. and 3) By learning and contrasting the user's intents under global and local views, MIDGN clearly captures the user's intents.
    \item \textbf{Graph models are effective in personalized bundle recommendation.} 
    Graph models (GCN, NGCF, RGCN) are effective in bundle recommendation, which can be proved by their better performance than MFBPR. We attribute this to their superiority in capturing graph structure and multi-hop collaborative information. Among these models, RGCN performs the best, which denotes the importance of distinguishing the different relationships for the problem of personalized bundle recommendation.
    \item \textbf{Capturing the user's preference on items' associations is important.} 
     Failing to capture bundle-level association, graph models (GCN, NGCF, RGCN) can not even surpass DAM, which employs attention mechanism and multi-tasks framework. By being devised to capture the user's preference on items' associations, BasConv and BGCN significantly improve the performance of personalized bundle recommendation and performs the best among baselines. 
\end{itemize}

\subsection{Study of MIDGN (RQ2$\textbf{\&}$RQ3)}   
Next we investigate the underlining mechanism of our MIDGN with three ablated models: $\text{MIDGN}_{\wo contra.}$ that removes the contrast module, $\text{MIDGN}_{\wo local}$ that replaces the graph disentangling module in the local view with GCN, and $\text{MIDGN}_{\wo global}$ that removes the graph disentangling module in the global view and uses GCN instead. 
From the results in \ref{tab:ablation}, we have the following observations:
\begin{itemize}
    \item MIDGN outperforms $\text{MIDGN}_{\wo global}$  significantly. This demonstrate the importance of global view in modeling the user's multiple intents. And involving more items is helpful to learn the user's intents.
    \item $\text{MIDGN}_{\wo local}$ is the least competitive. We contribute this to its failure on learning associations between items under the user's each intent. 
    \item MIDGN outperforms $\text{MIDGN}_{\wo contra.}$. This demonstrates the effectiveness of the intent contrast module in helping the intents from different views complement each other.
    \item All the ablated models significantly outperform the baselines. This demonstrates the effectiveness of disentangling multi-view intents for bundle recommendation.
\end{itemize}

{
\begin{table}[h]
 \setlength{\tabcolsep}{5pt}
 \centering
 \small
  {
    \begin{tabularx}{0.45\textwidth}{p{2cm}lXXlXX}
    \toprule
    \multirow{3}{*}{\text{Method}}&&\multicolumn{2}{c}{\text{NetEase}}&&\multicolumn{2}{c}{\text{Youshu}}\cr
    &&\multicolumn{2}{c}{\text{Metrics@20}}&&\multicolumn{2}{c}{\text{Metrics@20}}\cr
    \cline{3-4}  \cline{6-7}
    &&\text{Recall}&\text{NDCG}&&\text{Recall}&\text{NDCG}\cr
    \hline
    $\text{MIDGN}_{\wo     local}$&&0.0547&0.0287&&0.2475&0.1393\cr
    $\text{MIDGN}_{\wo  global}$&&0.0583&0.0303&&0.2464&0.1441\cr
     $\text{MIDGN}_{\wo  contra.}$&&0.0609&0.0312&&0.2588&0.1471\cr
    \hline
    
    MIDGN&&0.0678	&0.0343&&0.2682&0.1527\cr

    \hline
   
    \bottomrule
    \end{tabularx}
    }
    \caption{Ablated models analysis}
    \label{tab:ablation}
\end{table} 
}

{
\begin{table}[ht]
 \setlength{\tabcolsep}{5pt}
 \centering
 \small
    {
    \begin{tabularx}{0.45\textwidth}{p{1.5cm}lXXlXX}
    \toprule
    \multirow{3}{*}{\text{Method-L}}&&\multicolumn{2}{c}{\text{NetEase}}&&\multicolumn{2}{c}{\text{Youshu}}\cr
    &&\multicolumn{2}{c}{\text{Metrics@20}}&&\multicolumn{2}{c}{\text{Metrics@20}}\cr
    \cline{3-4}  \cline{6-7}
    &&\makecell[c]{\text{Recall}}&\makecell[c]{\text{NDCG}}&&\makecell[c]{\text{Recall}}&\makecell[c]{\text{NDCG}}\cr
    \hline
    MIDGN-1&&\makecell[c]{0.0442}&\makecell[c]{0.0233}&&\makecell[c]{0.2348}&\makecell[c]{0.1348}\cr
    
    MIDGN-2&&\makecell[c]{0.0559}&\makecell[c]{0.0295}&&\makecell[c]{0.2503}&\makecell[c]{0.1429}\cr
    
    MIDGN-3&&\makecell[c]{0.0642}&\makecell[c]{0.0331}&&\makecell[c]{0.2682}&\makecell[c]{0.1527}\cr
    
    MIDGN-4&&\makecell[c]{0.0678}&\makecell[c]{0.0343}&&\makecell[c]{0.2617}&\makecell[c]{0.1472}\cr
   
    \hline
   
    \bottomrule
    \end{tabularx}
    }
    \caption{Impact of Layer Number(L)}
    \label{tab:layer}
    \begin{tablenotes}
    \item
	\end{tablenotes}
\end{table} 
}

\begin{figure}[!ht]
    \includegraphics[width=0.45\textwidth]{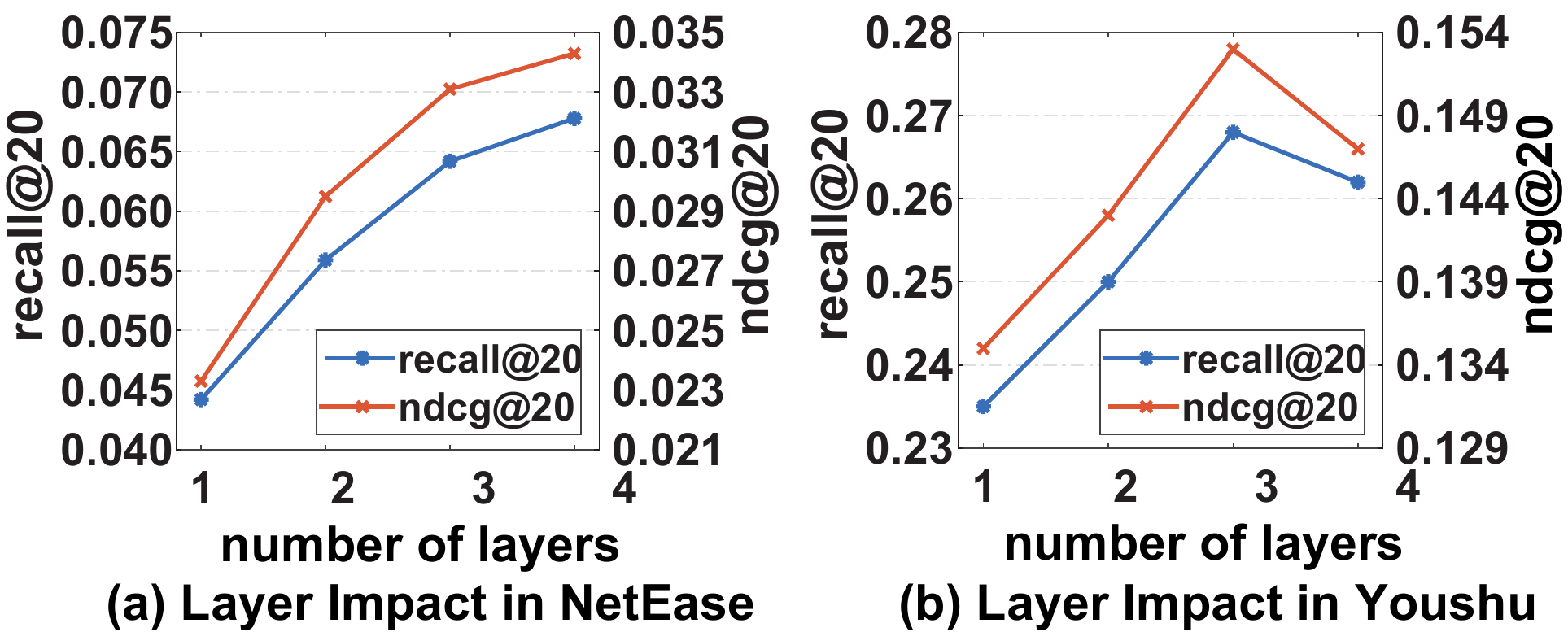}
    \caption{Impact of Layer Number(L)}
    \label{fig:layers}
  \end{figure}
  
We also conduct parameter studies to further investigate the influence of some parameters. Specifically, how the number of disentangling layers, the number of the user's intents influence the performance of MIDGN.

\textbf{Impact of Layer Number} The graph disentangling layer disentangles the user's intents and collects groups of items pertinent to individual intents to learn intent-aware representations. By stacking more layers, collaborative information from multi-hop neighbors is distilled. We investigate how the number of layers influences the performance of MIDGN. Specifically, we conduct experiments with layer $L$ in range $\{1, 2, 3, 4\}$ and the results are shown in Table \ref{tab:layer}, where MIDGN-L is to illustrate that L layers are involved in disentangling module. There are some observations:

\begin{itemize}
    \item Increasing the number of layers can improve the performance of our model. MIDGN-2 highly outperforms MIDGN-1. The reason is MIDGN-1 only gains information from one-hop neighbor and neglects high-order collaborative information. 
    \item When increasing the layer of number, the performance does not always improve. MIDGN-3 outperforms MIDGN-4 on data Youshu. This can be attributed to the noise which increases along with the hop of neighbor.
\end{itemize}

{
\begin{table}[ht]
\setlength{\tabcolsep}{5pt}
 \centering
 \small
    {
    \begin{tabularx}{0.45\textwidth}{p{1.5cm}lXXlXX}
    \toprule
    \multirow{3}{*}{\text{Method-K}}&&\multicolumn{2}{c}{\text{NetEase}}&&\multicolumn{2}{c}{\text{Youshu}}\cr
    &&\multicolumn{2}{c}{\text{Metrics@20}}&&\multicolumn{2}{c}{\text{Metrics@20}}\cr
    \cline{3-4}  \cline{6-7}
    &&\makecell[c]{\text{Recall}}&\makecell[c]{\text{NDCG}}&&\makecell[c]{\text{Recall}}&\makecell[c]{\text{NDCG}}\cr
    \hline
    MIDGN-1&&\makecell[c]{0.0579}&\makecell[c]{0.0299}&&\makecell[c]{0.2523}&\makecell[c]{0.1442}\cr
    
    MIDGN-2&&\makecell[c]{0.0652}	&\makecell[c]{0.0340}&&\makecell[c]{0.2632}&\makecell[c]{0.1497}\cr
    
    MIDGN-4&&\makecell[c]{0.0678}	&\makecell[c]{0.0343}&&\makecell[c]{0.2682}&\makecell[c]{0.1527}\cr
    
    MIDGN-8&&\makecell[c]{0.0587}	&\makecell[c]{0.0302}&&\makecell[c]{0.2517}&\makecell[c]{0.1439}\cr
   
    \hline
   
    \bottomrule
    \end{tabularx}
    }
    \caption{Impact of Intent Number(K)}
  \label{tab:Intent}
\end{table} 
}
\begin{figure}[!ht]
    \includegraphics[width=0.45\textwidth]{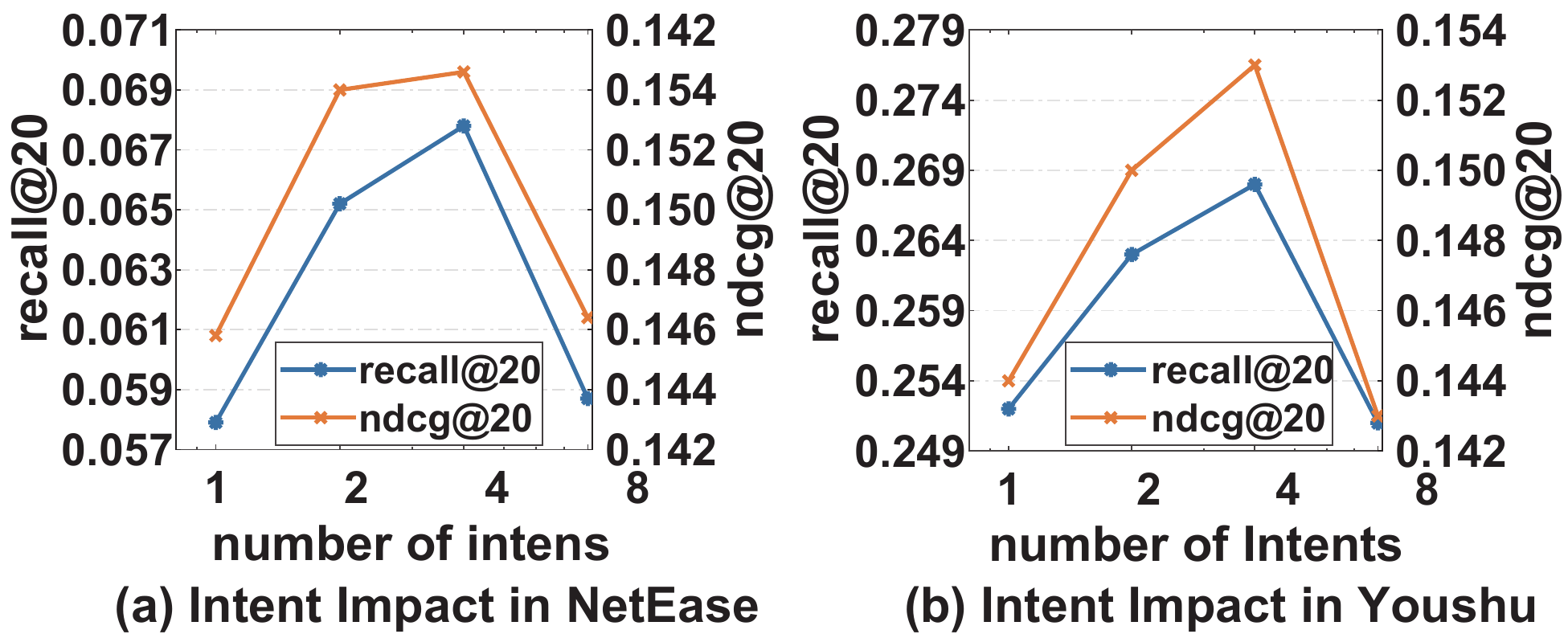}
    \caption{Impact of Intent Number(K)}
    \label{fig:Intent}
  \end{figure}
  
\textbf{Impact of Intent Number} To investigate the impact of intent number, we search the number $K$ of Intent in range $\{1, 2, 4, 8\}$. From the results in Table \ref{tab:Intent}, we have the following observations:
\begin{itemize}
    \item MIDGN performs the worst when the intent number $K=1$. This illustrates the diversity of the user's intents and an unitary intent can not present the user's preference effectively.
    \item When the intent number $K$ increases from $4$ to $8$, the performance drops sharply. This suggests the model suffers from too fine-grained intents. The reason may be that when cluster items into too many groups for individual intents, each group can not present the intent coupling with it well.
\end{itemize}
\section{Conclusion}
In this paper, we explore the diversity of the user's intents in the problem of personalized bundle recommendation. We propose a novel model named Multi-view Intent Disentangle Graph Networks(MIDGN) which disentangles the user's intents from both the global and local views. With the help of GNN equipped with the neighbor routing mechanism, MIDGN disentangles user-item and bundle-item graph coupling with the user's intents from global and local views, respectively. Meanwhile, MIDGN compares the intents learned from different views to better represent the user's preference as well as the items' associations at a finer grain of the user's intent.  Extensive experiments on Yoush and NetEase demonstrate the superiority of the proposed MIDGN.

\section*{Acknowledgments}
This work was supported in part by the National Natural Science Foundation of China under Grant No.61602197, Grant No.L1924068, Grant No.61772076, in part by CCF-AFSG Research Fund under Grant No.RF20210005, and in part by the fund of Joint Laboratory of HUST and Pingan Property \& Casualty Research (HPL). The authors would also like to thank the anonymous reviewers for their comments on improving the quality of this paper. 
\bibliography{paper}

\begin{thebibliography}{34}
\providecommand{\natexlab}[1]{#1}

\bibitem[{Bai et~al.(2019)Bai, Zhou, Song, Qu, An, Li, and
  Gao}]{bai2019personalized}
Bai, J.; Zhou, C.; Song, J.; Qu, X.; An, W.; Li, Z.; and Gao, J. 2019.
\newblock Personalized bundle list recommendation.
\newblock In \emph{The World Wide Web Conference}, 60--71.

\bibitem[{Berg, Kipf, and Welling(2017)}]{berg2017graph}
Berg, R. v.~d.; Kipf, T.~N.; and Welling, M. 2017.
\newblock Graph convolutional matrix completion.
\newblock \emph{arXiv preprint arXiv:1706.02263}.

\bibitem[{Cao et~al.(2017)Cao, Nie, He, Wei, Zhu, and Chua}]{cao2017embedding}
Cao, D.; Nie, L.; He, X.; Wei, X.; Zhu, S.; and Chua, T.-S. 2017.
\newblock Embedding factorization models for jointly recommending items and
  user generated lists.
\newblock In \emph{Proceedings of the 40th International ACM SIGIR Conference
  on Research and Development in Information Retrieval}, 585--594.

\bibitem[{Chang et~al.(2020)Chang, Gao, He, Jin, and Li}]{chang2020bundle}
Chang, J.; Gao, C.; He, X.; Jin, D.; and Li, Y. 2020.
\newblock Bundle recommendation with graph convolutional networks.
\newblock In \emph{Proceedings of the 43rd international ACM SIGIR conference
  on Research and development in Information Retrieval}, 1673--1676.

\bibitem[{Chen et~al.(2019)Chen, Liu, He, Gao, and Zheng}]{chen2019matching}
Chen, L.; Liu, Y.; He, X.; Gao, L.; and Zheng, Z. 2019.
\newblock Matching User with Item Set: Collaborative Bundle Recommendation with
  Deep Attention Network.
\newblock In \emph{IJCAI}, 2095--2101.

\bibitem[{Chen et~al.(2005)Chen, Tang, Shen, and Hu}]{chen2005market}
Chen, Y.-L.; Tang, K.; Shen, R.-J.; and Hu, Y.-H. 2005.
\newblock Market basket analysis in a multiple store environment.
\newblock \emph{Decision support systems}, 40(2): 339--354.

\bibitem[{Garfinkel et~al.(2006)Garfinkel, Gopal, Tripathi, and
  Yin}]{garfinkel2006design}
Garfinkel, R.; Gopal, R.; Tripathi, A.; and Yin, F. 2006.
\newblock Design of a shopbot and recommender system for bundle purchases.
\newblock \emph{Decision Support Systems}, 42(3): 1974--1986.

\bibitem[{He et~al.(2020)He, Deng, Wang, Li, Zhang, and Wang}]{he2020lightgcn}
He, X.; Deng, K.; Wang, X.; Li, Y.; Zhang, Y.; and Wang, M. 2020.
\newblock Lightgcn: Simplifying and powering graph convolution network for
  recommendation.
\newblock In \emph{Proceedings of the 43rd International ACM SIGIR conference
  on research and development in Information Retrieval}, 639--648.

\bibitem[{Kipf and Welling(2016)}]{kipf2016semi}
Kipf, T.~N.; and Welling, M. 2016.
\newblock Semi-supervised classification with graph convolutional networks.
\newblock \emph{arXiv preprint arXiv:1609.02907}.

\bibitem[{Le, Lauw, and Fang(2017)}]{le2017basket}
Le, D.~T.; Lauw, H.~W.; and Fang, Y. 2017.
\newblock Basket-sensitive personalized item recommendation.
\newblock IJCAI.

\bibitem[{Liu et~al.(2017)Liu, Fu, Chen, Xiong, and Chen}]{liu2017modeling}
Liu, G.; Fu, Y.; Chen, G.; Xiong, H.; and Chen, C. 2017.
\newblock Modeling buying motives for personalized product bundle
  recommendation.
\newblock \emph{ACM Transactions on Knowledge Discovery from Data (TKDD)},
  11(3): 1--26.

\bibitem[{Liu et~al.(2011)Liu, Ge, Li, Chen, and Xiong}]{liu2011personalized}
Liu, Q.; Ge, Y.; Li, Z.; Chen, E.; and Xiong, H. 2011.
\newblock Personalized travel package recommendation.
\newblock In \emph{2011 IEEE 11th International Conference on Data Mining},
  407--416. IEEE.

\bibitem[{Liu, Xie, and Lakshmanan(2014)}]{liu2014recommending}
Liu, Y.; Xie, M.; and Lakshmanan, L.~V. 2014.
\newblock Recommending user generated item lists.
\newblock In \emph{Proceedings of the 8th ACM Conference on Recommender
  systems}, 185--192.

\bibitem[{Liu et~al.(2020)Liu, Wan, Guo, Achan, and Yu}]{liu2020basconv}
Liu, Z.; Wan, M.; Guo, S.; Achan, K.; and Yu, P.~S. 2020.
\newblock Basconv: Aggregating heterogeneous interactions for basket
  recommendation with graph convolutional neural network.
\newblock In \emph{Proceedings of the 2020 SIAM International Conference on
  Data Mining}, 64--72. SIAM.

\bibitem[{Oord, Li, and Vinyals(2018)}]{oord2018representation}
Oord, A. v.~d.; Li, Y.; and Vinyals, O. 2018.
\newblock Representation learning with contrastive predictive coding.
\newblock \emph{arXiv preprint arXiv:1807.03748}.

\bibitem[{Parameswaran, Venetis, and
  Garcia-Molina(2011)}]{parameswaran2011recommendation}
Parameswaran, A.; Venetis, P.; and Garcia-Molina, H. 2011.
\newblock Recommendation systems with complex constraints: A course
  recommendation perspective.
\newblock \emph{ACM Transactions on Information Systems (TOIS)}, 29(4): 1--33.

\bibitem[{Rendle et~al.(2009)Rendle, Freudenthaler, Gantner, and
  Schmidt-Thieme}]{rendle2009bpr}
Rendle, S.; Freudenthaler, C.; Gantner, Z.; and Schmidt-Thieme, L. 2009.
\newblock BPR: Bayesian personalized ranking from implicit feedback.
\newblock In \emph{Proceedings of the Twenty-Fifth Conference on Uncertainty in
  Artificial Intelligence}, 452--461.

\bibitem[{Rendle et~al.(2012)Rendle, Freudenthaler, Gantner, and
  Schmidt-Thieme}]{rendle2012bpr}
Rendle, S.; Freudenthaler, C.; Gantner, Z.; and Schmidt-Thieme, L. 2012.
\newblock BPR: Bayesian personalized ranking from implicit feedback.
\newblock \emph{arXiv preprint arXiv:1205.2618}.

\bibitem[{Rendle et~al.(2014)Rendle, Freudenthaler, Gantner, and
  Schmidt-Thieme}]{rendle2014bayesian}
Rendle, S.; Freudenthaler, C.; Gantner, Z.; and Schmidt-Thieme, L.~B. 2014.
\newblock Bayesian personalized ranking from implicit feedback.
\newblock In \emph{Proc. of Uncertainty in Artificial Intelligence}, 452--461.

\bibitem[{Sar~Shalom et~al.(2016)Sar~Shalom, Koenigstein, Paquet, and
  Vanchinathan}]{sar2016beyond}
Sar~Shalom, O.; Koenigstein, N.; Paquet, U.; and Vanchinathan, H.~P. 2016.
\newblock Beyond collaborative filtering: The list recommendation problem.
\newblock In \emph{Proceedings of the 25th international conference on world
  wide web}, 63--72.

\bibitem[{Schlichtkrull et~al.(2018)Schlichtkrull, Kipf, Bloem, Van Den~Berg,
  Titov, and Welling}]{schlichtkrull2018modeling}
Schlichtkrull, M.; Kipf, T.~N.; Bloem, P.; Van Den~Berg, R.; Titov, I.; and
  Welling, M. 2018.
\newblock Modeling relational data with graph convolutional networks.
\newblock In \emph{European semantic web conference}, 593--607. Springer.

\bibitem[{Wang et~al.(2019)Wang, He, Wang, Feng, and Chua}]{wang2019neural}
Wang, X.; He, X.; Wang, M.; Feng, F.; and Chua, T.-S. 2019.
\newblock Neural graph collaborative filtering.
\newblock In \emph{Proceedings of the 42nd international ACM SIGIR conference
  on Research and development in Information Retrieval}, 165--174.

\bibitem[{Wang et~al.(2020{\natexlab{a}})Wang, Jin, Zhang, He, Xu, and
  Chua}]{wang2020disentangled}
Wang, X.; Jin, H.; Zhang, A.; He, X.; Xu, T.; and Chua, T.-S.
  2020{\natexlab{a}}.
\newblock Disentangled graph collaborative filtering.
\newblock In \emph{Proceedings of the 43rd International ACM SIGIR Conference
  on Research and Development in Information Retrieval}, 1001--1010.

\bibitem[{Wang et~al.(2020{\natexlab{b}})Wang, Tang, Lei, Song, Wang, and
  Zhang}]{wang2020disenhan}
Wang, Y.; Tang, S.; Lei, Y.; Song, W.; Wang, S.; and Zhang, M.
  2020{\natexlab{b}}.
\newblock DisenHAN: Disentangled Heterogeneous Graph Attention Network for
  Recommendation.
\newblock In \emph{Proceedings of the 29th ACM International Conference on
  Information \& Knowledge Management}, 1605--1614.

\bibitem[{Wang et~al.(2020{\natexlab{c}})Wang, Wei, Cong, Li, Mao, and
  Qiu}]{wang2020global}
Wang, Z.; Wei, W.; Cong, G.; Li, X.-L.; Mao, X.-L.; and Qiu, M.
  2020{\natexlab{c}}.
\newblock Global context enhanced graph neural networks for session-based
  recommendation.
\newblock In \emph{Proceedings of the 43rd International ACM SIGIR Conference
  on Research and Development in Information Retrieval}, 169--178.

\bibitem[{Wang et~al.(2020{\natexlab{d}})Wang, Wei, Cong, Li, Mao, Qiu, and
  Feng}]{wang2020exploring}
Wang, Z.; Wei, W.; Cong, G.; Li, X.-L.; Mao, X.-L.; Qiu, M.; and Feng, S.
  2020{\natexlab{d}}.
\newblock Exploring Global Information for Session-based Recommendation.
\newblock \emph{arXiv preprint arXiv:2011.10173}.

\bibitem[{Wang et~al.(2020{\natexlab{e}})Wang, Wei, Cong, Mao, Li, and
  Feng}]{wang2020exploiting}
Wang, Z.; Wei, W.; Cong, G.; Mao, X.-L.; Li, X.-L.; and Feng, S.
  2020{\natexlab{e}}.
\newblock Exploiting Repeated Behavior Pattern and Long-term Item dependency
  for Session-based Recommendation.
\newblock \emph{arXiv preprint arXiv:2012.05422}.

\bibitem[{Wei et~al.(2019)Wei, Liu, Mao, Guo, Zhu, Zhou, and
  Hu}]{wei2019emotion}
Wei, W.; Liu, J.; Mao, X.; Guo, G.; Zhu, F.; Zhou, P.; and Hu, Y. 2019.
\newblock Emotion-aware chat machine: Automatic emotional response generation
  for human-like emotional interaction.
\newblock In \emph{Proceedings of the 28th ACM International Conference on
  Information and Knowledge Management}, 1401--1410.

\bibitem[{Wu et~al.(2019)Wu, Tang, Zhu, Wang, Xie, and Tan}]{wu2019session}
Wu, S.; Tang, Y.; Zhu, Y.; Wang, L.; Xie, X.; and Tan, T. 2019.
\newblock Session-based recommendation with graph neural networks.
\newblock In \emph{Proceedings of the AAAI Conference on Artificial
  Intelligence}, volume~33, 346--353.

\bibitem[{Xie, Lakshmanan, and Wood(2010)}]{xie2010breaking}
Xie, M.; Lakshmanan, L.~V.; and Wood, P.~T. 2010.
\newblock Breaking out of the box of recommendations: from items to packages.
\newblock In \emph{Proceedings of the fourth ACM conference on Recommender
  systems}, 151--158.

\bibitem[{Xie, Lakshmanan, and Wood(2014)}]{xie2014generating}
Xie, M.; Lakshmanan, L.~V.; and Wood, P.~T. 2014.
\newblock Generating top-k packages via preference elicitation.
\newblock \emph{Proceedings of the VLDB Endowment}, 7(14): 1941--1952.

\bibitem[{Zhao et~al.(2019)Zhao, Zhang, Wang, Zhou, and
  Cheng}]{zhao2019recurrent}
Zhao, S.; Zhang, Y.; Wang, S.; Zhou, B.; and Cheng, C. 2019.
\newblock A recurrent neural network approach for remaining useful life
  prediction utilizing a novel trend features construction method.
\newblock \emph{Measurement}, 146: 279--288.

\bibitem[{Zheng et~al.(2010)Zheng, Cao, Zheng, Xie, and
  Yang}]{zheng2010collaborative}
Zheng, V.; Cao, B.; Zheng, Y.; Xie, X.; and Yang, Q. 2010.
\newblock Collaborative filtering meets mobile recommendation: A user-centered
  approach.
\newblock In \emph{Proceedings of the AAAI Conference on Artificial
  Intelligence}, volume~24.

\bibitem[{Zhu et~al.(2014)Zhu, Harrington, Li, and Tang}]{zhu2014bundle}
Zhu, T.; Harrington, P.; Li, J.; and Tang, L. 2014.
\newblock Bundle recommendation in ecommerce.
\newblock In \emph{Proceedings of the 37th international ACM SIGIR conference
  on Research \& development in information retrieval}, 657--666.

\end{thebibliography}

\end{document}